\documentclass[floatfix,twocolumn,pre,superscriptaddress]{revtex4}
\usepackage{graphicx}
\usepackage{graphics}
\usepackage{latexsym}
\usepackage{amsmath, amsthm, amssymb, wasysym}
\usepackage{xcolor}
\usepackage{epsfig}

\begin{document}

\title{Transport and tumbling of polymers in viscoelastic shear flow}
\author{Sadhana Singh}
\affiliation{Department of Physics, Banaras Hindu University, Varanasi-221005}
\author{R. K. Singh}
\email[]{rksinghmp@gmail.com}
\affiliation{Department of Physics, Banaras Hindu University, Varanasi-221005}
\affiliation{Department of Physics, Indian Institute of Technology, Bombay, Powai,
Mumbai-400076}
\author{Sanjay Kumar}
\affiliation{Department of Physics, Banaras Hindu University, Varanasi-221005}

\begin{abstract}
Polymers in shear flow are ubiquitous and we study their motion in a viscoelastic
fluid under shear. Employing dumbbells as representative, we find that the center
of mass motion follows: $\langle x^2_c(t) \rangle \sim \dot{\gamma}^2
t^{\alpha+2}, ~0< \alpha <1$, generalizing the earlier result: $\langle x^2_c(t) \rangle
\sim \dot{\gamma}^2t^3 ~(\alpha = 1)$. Motion of the relative coordinate, on the other hand,
is quite
intriguing in that $\langle x^2_r(t) \rangle \sim t^\beta$ with $\beta = 2(1-\alpha)$
for small $\alpha$. This implies nonexistence of the steady state. We remedy this
pathology by introducing a nonlinear spring with FENE-LJ interaction and study
tumbling dynamics of the dumbbell. The overall effect of viscoelasticity is to slow down
the dynamics in the experimentally observed ranges of the Weissenberg number.
We numerically obtain the characteristic time of tumbling and show that small
changes in $\alpha$ result in large changes in tumbling times.
\end{abstract}

\maketitle

\textit{Introduction}:
Viscoelastic fluids under shear are ubiquitous, especially in biological systems,
and aid in transport of biomolecules \textit{in-vivo}. Viscoelasticity, as the name
suggests, is the property of a material comprising of both viscous and elastic behavior
\cite{maxwell}. Almost all materials with biological or engineering interests
are viscoelastic to some degree \cite{lakes}. The elastic component of the material tends to
bring it back to its original configuration when put under stress \cite{jones}.
As a result, motion in viscoelastic media is generally slower,
\textit{i.e.}- the mean square displacement
$\langle x^2(t) \rangle \sim t^\alpha$ \cite{jeon}, with $0 < \alpha < 1$,
consequent of the anti-persistent
correlations in successive displacements \cite{mandelbrot}.
Viscoelastic subdiffusion frequently arises in motion in biological domains,
\textit{e.g.}- motion in
crowded fluids \cite{Mweiss}, cytoplasm of living cells \cite{weber}, locus of a
chromosome in eukaryotes \cite{israel}, etc.

Even though a useful representative of system dynamics, a single particle description
is not fully appropriate when it comes to investigating systems with internal degrees
of freedom, \textit{e.g.}- polymers. In addition, polymers constitute the basic
building blocks of the macromolecules like DNA and proteins. Hence, it becomes natural
to investigate the dynamical aspects of a polymer in viscoelastic media.
However, most of the polymer transport
\textit{in-vivo} takes place in viscoelastic fluids under shear, wherein
they not only move but also tumble along, \textit{i.e.}- an end-to-end
rotation. The phenomena of polymer tumbling is well understood for the case of
viscous shear flows \cite{smith, leduc}. And arises when the relaxation time of
the polymer is larger than the time-scale of flow deformation \cite{usabiaga}, with
characteristic tumbling time varying sublinearly with the flow rate \cite{schroeder,
winkler}. However, a majority of studies involving tumbling
do not cover the practically important case of shear flows arising in viscoelastic
media, \textit{e.g.}- polymer plastics and most of the biological materials
\cite{ozkaya}. 

These observations raise an interesting question: what are the dynamical characteristics of
a polymer in a viscoelastic fluid under shear? This is a question of immense
practical significance, which we answer in the present work employing a dumbbell
which is the simplest form of a polymer.
For the two masses connected by a harmonic spring, we show both analytically and
numerically that the separation grows without bounds. This implies towards the nonexistence
of steady state and essentially means that tumbling cannot be addressed using
a linear system. We remedy this pathology by introducing a finitely extensible nonlinear
elastic spring with repulsive part of the Lennard-Jones interaction (FENE-LJ) \cite{herrchen,
grest}. Thus, allowing us to address tumbling.

\textit{Generalized Langevin equation in shear flows}:
The generalized Langevin equation (GLE) \cite{zwanzig} describing the motion of
a dumbbell in a viscoelastic material under shear reads:
\begin{align}
\label{gle}
\int^t_0 dt' \eta(t-t')(\dot{\bf r}_i-\dot{\gamma}y_i{\bf i})(t') 
&= -\nabla_i V(|{\bf r}_i - {\bf r}_j|) + 
{\bf \xi}_i(t),
\end{align}
where ${\bf r}_i \equiv (x_i, y_i, z_i)$,
with $i=1,2$ and $i\neq j$ denote the two particles.
The shear rate $\dot{\gamma}$ defines the Weissenberg number Wi $=\tau_0 
\dot{\gamma}$, in terms of the relaxation time $\tau_0$ of the dumbbell in
the absence of any flow.
The noise vectors ${\bf \xi}_1$ and ${\bf \xi}_2$ are Gaussian random 
variables with correlation matrices:
$\langle {\bf \xi}_i(t) {\bf \xi}^T_j(t') \rangle = \delta_{ij} k_B T \eta(|t-t'|){\bf I}_3$,
consistent with the fluctuation dissipation relation \cite{kubo}, where ${\bf I}_3$
denotes the $3\times 3$ identity matrix.
For the case of harmonically interacting dumbbells we choose $V(|{\bf r}_i - {\bf r}_j|) =
\frac{1}{2}\omega^2_0({\bf r}_i - {\bf r}_j)^2$, which is a Rouse polymer of size $N=2$
\cite{rouse}.

The term inside the integral in Eq.~(\ref{gle}) is the memory kernel representing 
a time-dependent friction. Consequently, the present state depends on the
entire history. Physically, the GLE renders itself derivable in terms of
mechanical equations for a particle interacting with a thermal bath, in terms of the
spectral density of the bath  oscillators \cite{weiss, raz}. To address the problem at
hand, we employ a power law decaying form for the memory kernel:
$\eta(t) = \eta_\alpha t^{-\alpha}/\Gamma(1-\alpha)$, with $0 < \alpha < 1$ \cite{gemant}.
With this form of memory kernel,
the GLE results in a subdiffusion for the motion of a free particle \cite{goychuk, wang}.

\textit{Center of mass motion for linear spring}:
Absence of any external force on the dumbbell allows 
us to separate its dynamics into the motion of center of mass and motion about 
the center of mass. The coordinate of the center of mass ${\bf r}_c = \frac{{\bf r}_1+
{\bf r}_2}{2}$ evolves as:
\begin{align}
\label{gle_cm}
\int^t_0 dt' \eta(t-t')(\dot{\bf r}_c-\dot{\gamma}y_c{\bf i})(t') = {\bf \xi}_c(t),
\end{align}
where $\xi_c(t)=[\xi_1(t)+\xi_2(t)]/2$ is Gaussian noise with mean zero and 
correlation:
$\langle \xi_c(t)\xi^T_c(t') \rangle = \frac{1}{2}k_BT \eta(|t-t'|){\bf I}_3$.
\begin{figure}
\includegraphics[width=0.5\textwidth]{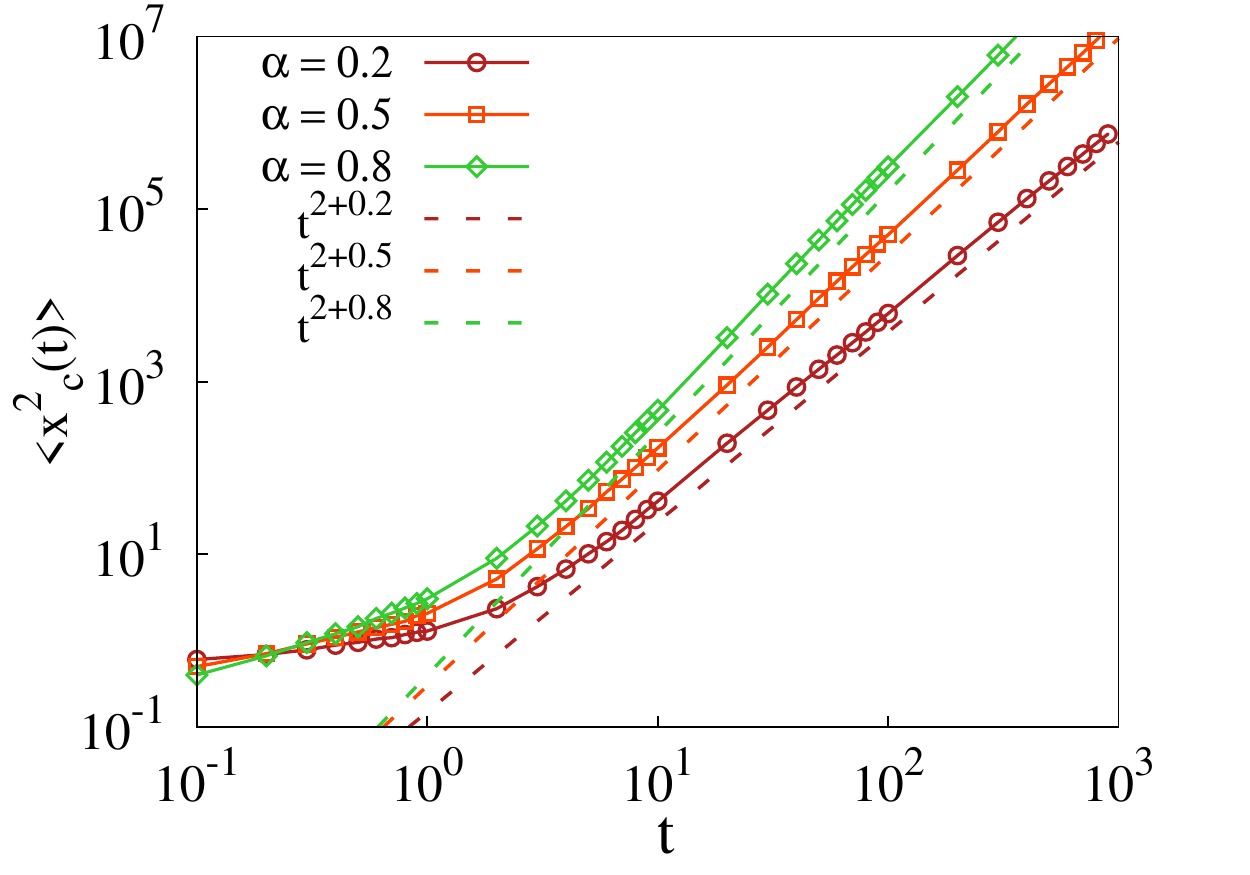}
\caption{Mean square displacement of the center of mass motion $x_c$ along the shear
flow direction for varying $\alpha$, with respective line fits. Parameter values for
numerical calculation are: $\dot{\gamma} = 1,~\eta_\alpha = 1$ and $k_BT/2=1$.}
\label{fig1}
\end{figure}
It is evident looking at Eq.~(\ref{gle_cm}) that the center of mass 
moves like a free particle in shear flow. Interestingly, the $y$ (and $z$) components
of motion do not feel the effect of shear flow, with the well known two-point correlation:
$\langle y(t_1)y(t_2) \rangle = k_BT/2\eta_\alpha[t_1^\alpha + t_2^\alpha -
|t_1-t_2|^\alpha$] \cite{jeon}. The $x$ component of motion is, however, affected
by the presence of shear flow which is directed along the $x$-axis. 
Invoking the Laplace transform of Eq.~(\ref{gle_cm}) allows us to decouple the
convolution of the memory kernel $\eta$ and local velocity. As a result, the
time evolution of the $x$-coordinate of the center of mass evolves as:
\begin{align}
x_c(t) = \dot{\gamma}\int^t_0 dt' y_c(t') + \int^t_0 dt' g(t-t')\xi_{cx}(t'),
\end{align}
where $\tilde{g}(s) = 1/s\tilde{\eta}(s)$, is the Laplace transform of $g$.
This allows us to address the effect of shear flow on center of mass motion, which in
terms of mean square displacement reads:
\begin{align}
\langle x^2_c(t) \rangle &= \frac{\dot{\gamma}^2 k_BT}{\eta_\alpha}\frac{\alpha+1}
{\Gamma(\alpha+3)}t^{\alpha+2} + \frac{k_BT}{\eta_\alpha}\frac{t^\alpha}{\Gamma(\alpha+1)},
\nonumber\\
&\approx \frac{\dot{\gamma}^2 k_BT}{\eta_\alpha}\frac{\alpha+1}
{\Gamma(\alpha+3)}t^{\alpha+2},
\end{align}
with the shear contribution dominating at large times. This is an interesting result,
in that motion along the flow is shear
dominated and thermal fluctuations play only a sub-dominant role. It also generalizes
the earlier study on viscous shear flows ($\alpha = 1$):
$ \langle x^2_c(t) \rangle = \frac{2}{3}\dot{\gamma}^2 D t^3$ \cite{mcphie},
with $D = k_BT/2\eta_\alpha$. Numerical solution of Eq.~(\ref{gle_cm})
provides a confirmation of our analytical results \cite{supp, advchemphys}(\textit{cf.}
Fig.~\ref{fig1}). The subdiffusive nature of motion at small times viz. $t \lessapprox 1$
is also discernible from Fig.~\ref{fig1}. This implies that the shear flow results in a
crossover from subdiffusive motion to a motion faster than ballistic.

\textit{Relative motion for linear spring}:
The relative coordinate ${\bf r}_r = {\bf r}_1-{\bf r}_2$ evolves as:
\begin{align}
\label{dxrdt}
\int^t_0 dt' \eta(t-t')(\dot{\bf r}_r-\dot{\gamma}y_r{\bf i})(t') = 
-2\omega^2_0{\bf r}_r + {\bf \xi}_r(t),
\end{align}
and represents a particle moving in a harmonic potential in a viscoelastic medium
under shear. The noise variable 
$\xi_r(t) = \xi_1(t)-\xi_2(t)$ is an unbiased colored Gaussian noise with
correlation matrix $\langle \xi_r(t) \xi^T_r(t') \rangle = 2k_B T\eta(|t-t'|){\bf I}_3$.

As the $y$ (and $z$) component does not feel the effect of flow, its dynamics is
known exactly: $\langle y_r(t)y_r(t') \rangle = 2k_BT[Q(t)+Q(t')-2\omega^2_0Q(t)Q(t')-
Q(|t-t'|)]$\cite{vinales, desposito},
where $Q(t) = \frac{1}{2\omega^2_0}[1-E_\alpha(-a t^\alpha)]$ \cite{desposito2009} where
$a = \frac{2\omega^2_0}{\eta_\alpha}$ and $E_\alpha()$ is the Mittag-Leffler function
\cite{podlubny}. The motion along $x$ direction, however, feels the effect of both thermal
fluctuations and shear flow, with the former known exactly. The shear contribution to
the motion in Laplace domain reads: $\tilde{x}_r(s)_{Sh} = \dot{\gamma}\tilde{G}(s)
\tilde{y}_r(s)$ with $\tilde{G}(s) = \frac{s^{\alpha-1}}{s^\alpha+a}$, and allows us to
calculate the fluctuations in the relative coordinate:
$\frac{\langle x^2_r(t) \rangle_{Sh}}{\dot{\gamma}^2} = \int^t_0 dt_1 G(t-t_1)
\int^t_0 dt_2 G(t-t_2)\langle y_r(t_1) y_r(t_2) \rangle$. Now, using the 2-point
correlation of $y$ and using the following integrals: $\int^t_0 dt_1 G(t-t_1) =
tE_{\alpha, 2}(-at^\alpha)$, $\int^t_0 dt_1 G(t-t_1)Q(t_1) = (G*Q)(t) = R(t) =
\frac{at^{\alpha+1}}{2\omega^2_0} E^{(1)}_{\alpha, 2}(-at^\alpha)$ and
$\int^t_0 dt_1 G(t-t_1)\int^t_0dt_2G(t-t_2)Q(|t_1-t_2|) = 2\int^t_0 dt_1 G(t_1)R(t_1)$,
we have:
\begin{align}
\label{xr_sq}
\frac{\langle x^2_r(t) \rangle_{Sh}}{2\dot{\gamma}^2k_BT} = 
\frac{t^2}{\omega^2_0}\sum^\infty_0\sum^\infty_0
\frac{l(-at^\alpha)^{k+l}}{\Gamma(\alpha k+1)\Gamma(\alpha l+2)[\alpha(k+l)+2]}\nonumber\\
+\frac{t^2}{2\omega^2_0}
[2at^\alpha E_{\alpha,2}(-at^\alpha)E^{(1)}_{\alpha, 2}(-at^\alpha)-
a^2t^{2\alpha}\{E^{(1)}_{\alpha, 2}(-at^\alpha)\}^2],
\end{align}
where $E_{\alpha,\beta}()$ is the two-parameter Mittag-Leffler function and $E^{(1)}_
{\alpha,\beta}()$ its derivative \cite{podlubny}. Let us make an approximation, viz.
\begin{figure}
\includegraphics[width=0.5\textwidth]{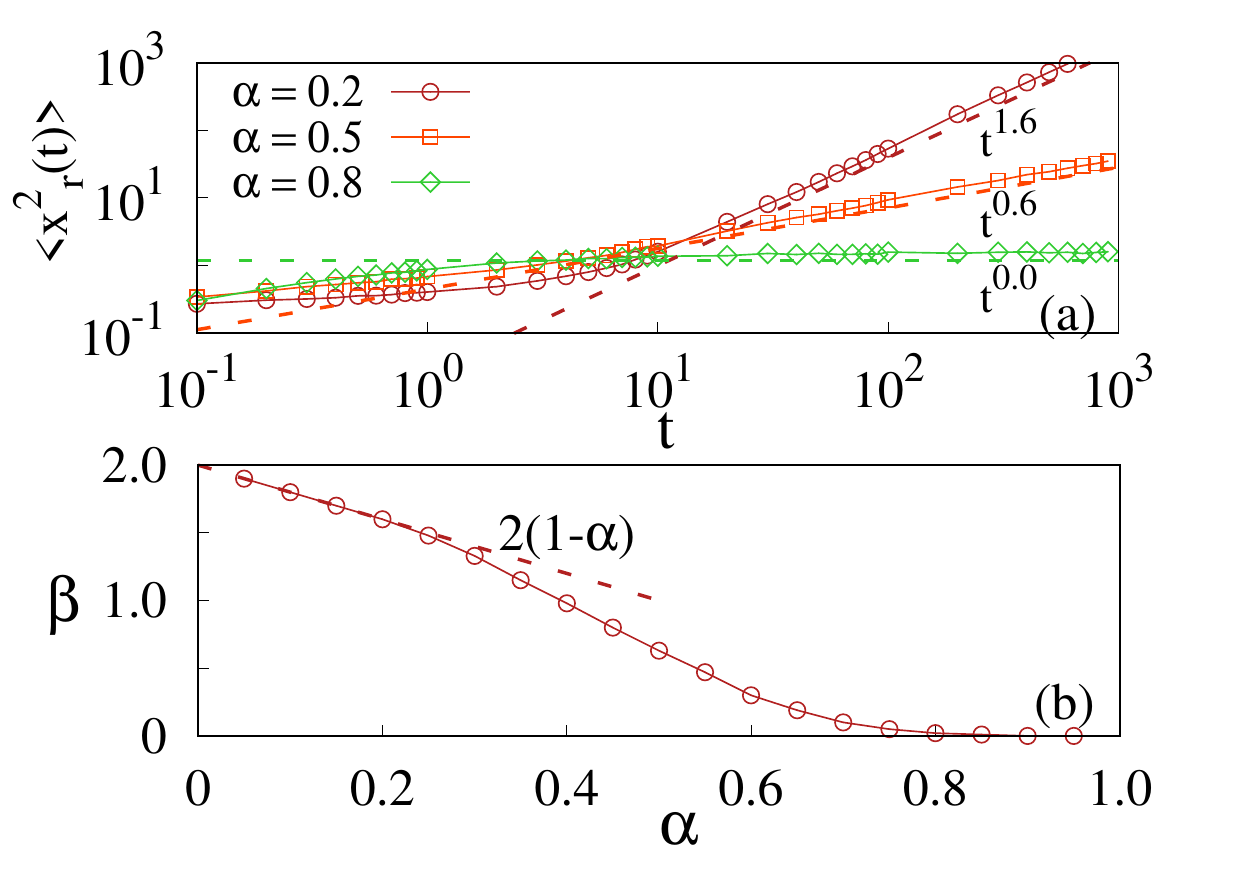}
\caption{(a) Mean square displacement of the relative motion $x_r$
along the shear flow direction for various $\alpha$, with respective line fits.
(b) Dependence of $\beta$ on the exponent of subdiffusion $\alpha$. Parameters for
the numerical calculation are: $\dot{\gamma} = 1,~\eta_\alpha = 1, ~2k_BT = 1$ and
$\omega_0=1$.}
\label{fig2}
\end{figure}
$\frac{l}{\alpha(k+l)+2} \approx \frac{l}{2}$, which is expected to hold for
small values of $\alpha$, for the first term in Eq.~(\ref{xr_sq}) and the asymptotic
forms of $E$ and $E^{(1)}$ \cite{regan} in the second term, we have:
\begin{align}
\label{xr_apr}
\langle x^2_r(t) \rangle_{Sh}
\sim \frac{\alpha\dot{\gamma}^2k_BT}{a^2\omega^2_0\Gamma^2(2-\alpha)}t^{2-2\alpha}.
\end{align}
This implies that the shear contribution to the motion of the relative coordinate grows
without bounds, in a superdiffusive manner ($\alpha$ is small). In addition, as the thermal
contribution eventually reaches a steady value, this is the fate of separation between the
two masses in that $\langle x^2_r(t) \rangle \sim t^\beta$ with $\beta = 2(1-\alpha)$ (for
small $\alpha$). For arbitrary values of
$\alpha$, such a closed form expression is not possible, and we solve Eq.~(\ref{dxrdt})
numerically to assess the behavior of fluctuations in the separation of the two masses.
We show the results for $\langle x^2_r(t) \rangle$ in Fig.~\ref{fig2}(a) for different
values of $\alpha$. For the entire range of $\alpha \in (0,1)$, Fig.~\ref{fig2}(b) shows
that the fluctuations in the relative coordinate go from superdiffusive to diffusive to
subdiffusive as $\alpha$ grows from 0 to 1\cite{supp}. The deviation from the straight
line behavior is also evident, implying towards the failure of the approximation made
to decouple the series in Eq.~(\ref{xr_sq}).

Nonexistence of the steady-state for the motion of relative coordinates implies that
the system does not feel the effects of confinement. Hence, the harmonically
interacting dumbbell which serves as a starting point for addressing tumbling in viscous
shear flows, \textit{e.g.}- Rouse chains \cite{das1, das2}, does not work for
motion in viscoelastic medium under shear. As a result, we need a potential strong enough
to exhibit a nonequilibrium steady state for the motion of relative coordinates.
In the next section, we address this problem by introducing nonlinear interactions
and study tumbling of dumbbells in viscoelastic shear flows.

\textit{Generalized Langevin equation in shear flows-Nonlinear dumbbell
model}: In order to bring in nonlinearity in the problem, we introduce FENE-LJ
potential which is more realistic compared to the harmonic interaction \cite{grest}.
The inter-particle interaction is a contribution
from both repulsive and attractive parts, viz. $V = V_{LJ} + V_{FENE}$, wherein
\begin{subequations}
\begin{align}
&V_{LJ}(r) = 4 \varepsilon [(\sigma/r)^{12}-(\sigma/r)^6],~\text{and}\\
&V_{FENE}(r) = -(kR^2_0/2)\ln[1-(r/R_0)^2].
\end{align}
\end{subequations}
As mentioned earlier, we consider only the repulsive part of $V_{LJ}$. In above equations,
$r = |{\bf r}_1-{\bf r}_2|$ denotes the separation
between the two monomers, $\sigma$ their size, $\varepsilon$ the strength of repulsion,
$R_0$ the maximum extension, and $k$ the stiffness constant.
The force on particle $i$ due to $j$ is $-\nabla_i V(|{\bf r}_i-{\bf r}_j|)$, with
$i, j = 1, 2$. The resulting equations of motion read:
\begin{align}
\label{in_gle}
\ddot{\bf r}_i + \int^t_0 dt' \eta(t-t')(\dot{\bf r}_i&-\dot{\gamma}y_i{\bf i})(t')\nonumber\\
&= -\nabla_i V(|{\bf r}_i - {\bf r}_j|) + \dot{\gamma}\dot{y}_i {\bf i} + {\bf \xi}_i(t),
\end{align}
with $i,~j=1, 2$ and $i\neq j$. We have retained the acceleration terms for the nonlinear
system because of its numerical advantage (avoids root finding like in the overdamped case)
\cite{supp}.
The second term on the right hand side of the above equations, $\dot{\gamma} \dot{y}_i$ 
is the coordinate dependent contribution from the 
flow. This term arises when we take account of the local streaming velocity alongwith 
the actual momentum \cite{mcphie}.
\begin{figure}
\includegraphics[width=0.5\textwidth]{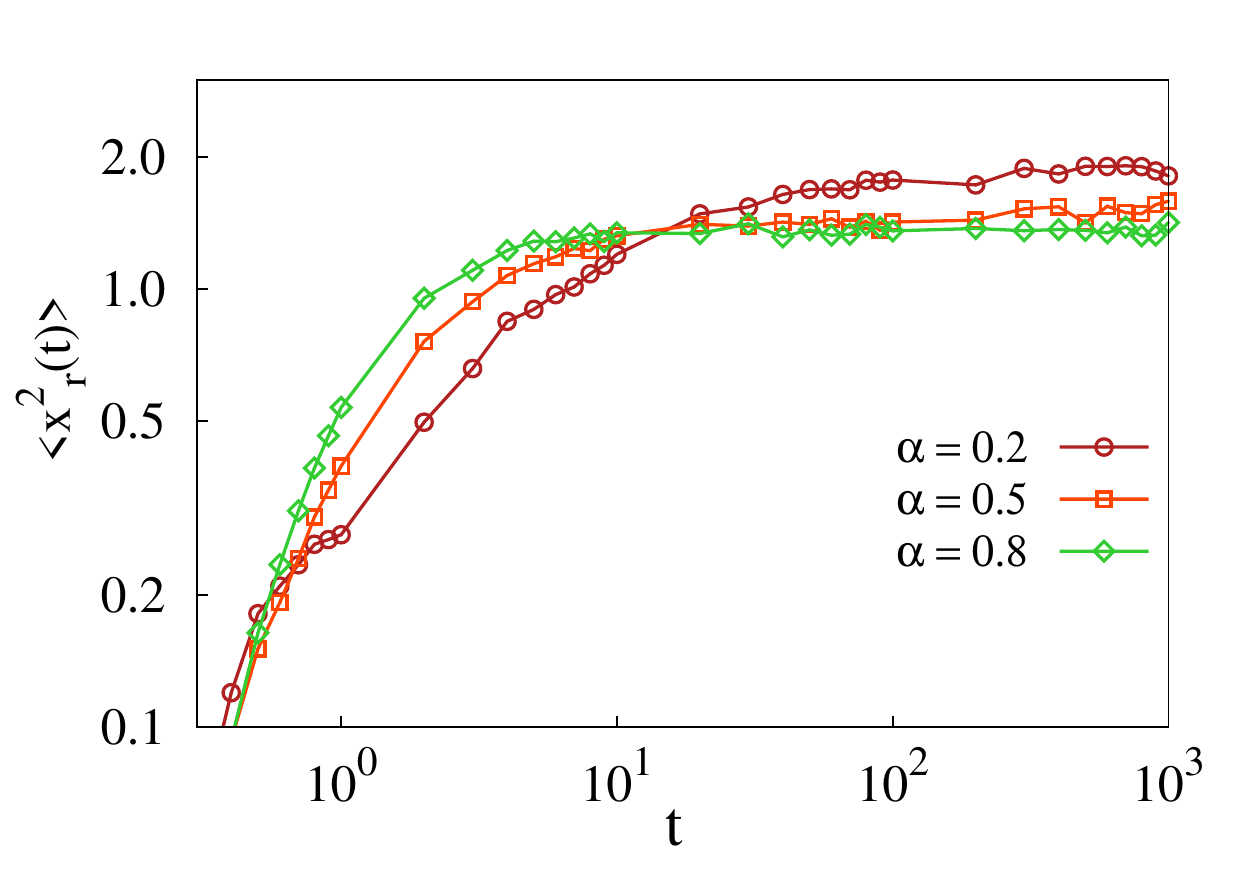}
\caption{Mean square displacement of relative separation $x_r$ of two particle connected
by a nonlinear spring for different values of $\alpha$.}
\label{fene_fig}
\end{figure}
Similar to the case of overdamped motion, the center of mass for underdamped motion
also follows $\langle x^2_c(t) \rangle \sim t^{\alpha+2}$. The nonlinear spring, however,
unlike the harmonic spring, achieves a steady-state owing the FENE-LJ potential which keeps
the bond-length in the interval $(\sigma, R_0)$. This is evident from the behavior of the
mean square displacement $\langle x^2_r(t) \rangle$ of the relative separation between
the two masses connected by the nonlinear spring (\textit{cf.} Fig.~\ref{fene_fig}).

In what follows, energy is measured in units of $\varepsilon$ and  distance in units
of $\sigma$. In addition, following \cite{kremer, sdhn}, we choose: $\varepsilon = 1$,
$\sigma = 1$, $R_0={1.5 \sigma}$, $k =30 \varepsilon/\sigma^2$,
$k_B T = 1.2\varepsilon$ and $\eta_\alpha = 7.5$.

\textit{Distribution of tumbling times}:
\begin{figure}
\includegraphics[width=0.5\textwidth]{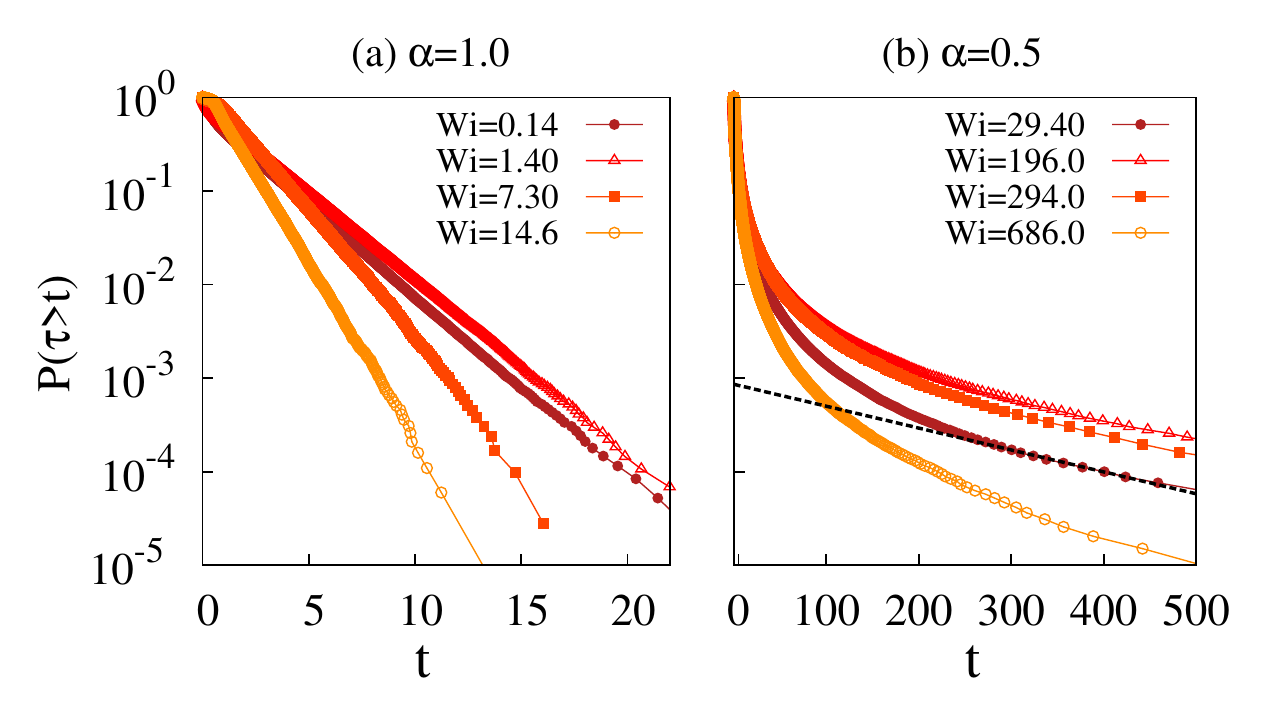}
\caption{Probability distribution $P(\tau \ge t)$ for the dumbbells tumbling in
(a) purely viscous shear flow ($\alpha = 1$) vs (b) viscoelastic shear flows
($\alpha = 0.5$). The exponentially decaying tails for the two cases are discernible
from the graphs (line fit for Wi$=29.40$), with the distributions of the form
$P(\tau \ge t) \approx \exp(-\nu\tau)$.
In addition, a change in the trend of slope for the two types of flows is also
clearly evident.}
\label{fig3}
\end{figure}
Tumbling time $\tau$ is defined as the interval of successive zero-crossings of
the end-to-end vector $R_x=x_1-x_2$. The distribution of tumbling times $P(\tau \ge t)$
in purely viscous flow exhibits exponentially decaying tails for various values
of the Weissenberg number Wi (Fig.~\ref{fig3} (a)). Interestingly, even for
tumbling in viscoelastic shear flow,
$P(\tau \ge t)$ exhibits exponentially decaying tails (\textit{cf.} Fig.~\ref{fig3} (b)),
though the time $\tau$ taken for the viscoelastic case is much longer when compared
to its viscous counterpart. In addition, the tumbling events for the two types
of flows, viz. viscous and viscoelastic case occur at different levels of flow strengths.
As a matter of fact, the observed values of flow strength for the viscoelastic case are
well beyond the experimentally observed ranges for the case of viscous shear flows.
This is evident from the respective values of Wi for the two cases, which are at least
an order of magnitude apart. The reason for this difference is that relaxation time in
a viscoelastic medium is much longer compared to relaxation in a purely viscous fluid.
In other words, the effect of viscoelasticity in the medium is to slow down the
characteristic tumbling frequency at finite Wi consistent with earlier studies
\cite{1,2,3} for rotational dynamics of suspended particle in viscoelastic shear flow.
The exponentially decaying tails of the tumbling time distribution,
$P(\tau \ge t) \approx \exp(-\nu\tau)$, define the characteristic tumbling time
$\tau_{tumb}$ as inverse of the characteristic exponent $\nu$. In addition, for
the case of purely viscous flows, \textit{i.e.}- for $\alpha = 1$, we find via
numerical calculations that $\nu \tau_0 \approx \text{Wi}^{0.67}$ for Wi~$\gg 1$
\cite{celani}, wherein $\tau_0$ is relaxation time of the autocorrelation
$\langle R_x(t)R_x(t+T) \rangle$.

\textit{Effect of subdiffusion on tumbling}:
\begin{figure}
\includegraphics[width=0.5\textwidth]{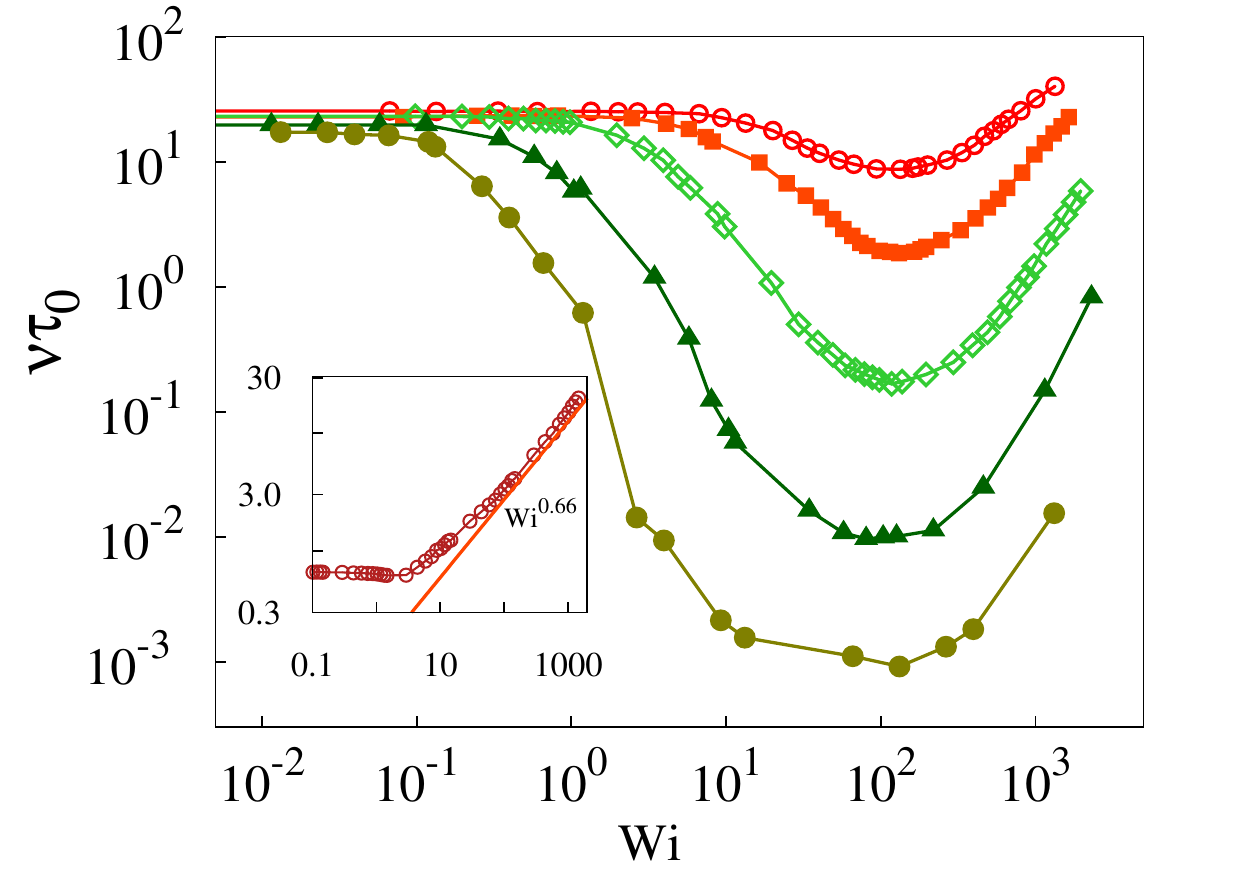}
\caption{Limiting decay rate $\nu$ scaled with $\tau_0$ vs Wi for subdiffusive case
for different values of $\alpha$. For weak flow strengths, \textit{i.e.} for
Wi~$\ll 1$, the different curves trace each other. On the other hand, for Wi~$\gg 1$,
the $\nu\tau_0$ grows with the Weissenberg number as Wi$^{\mu(\alpha)}$, where the
exponent $\mu$ depends on $\alpha$.
Inset shows the variation of dimensionless tumbling frequency $\nu\tau_0$ with the
Weissenberg number Wi for a dumbbell tumbling in a purely viscous medium ($\alpha = 1$).
}
\label{fig5}
\end{figure}
We study the effect of subdiffusion on tumbling of dumbbells in Fig.~\ref{fig5},
wherein we find that $\nu\tau_0$ exhibits a nonmonotonic behavior with Wi
which is absent in viscous medium (\textit{cf.} inset of Fig.~\ref{fig5}).
Though there is a slight deviation in the
trends of the curve for distribution with increasing Wi (\textit{cf.} Fig.~\ref{fig3}
(a) for $\alpha = 1$) showing nonmonotonicity in value of $\nu$ but it is negligible
compared to the case for $\alpha < 1$. To understand this behavior of $\nu\tau_0$ vs Wi,
let us have a look at Eq.~(\ref{in_gle}), from where it is clear that the relative
coordinate ${\bf r}_1-{\bf r}_2$ follows the motion of a single particle in the potential
$V=V_{FENE}+V_{LJ}$, and in the absence of shear
relaxes at its natural time scale dictated by the parameters of the potential
and the degree of subdiffusion $\alpha$. However, when a small but finite amount of
shear flow is introduced, it tends to align the dumbbell along the flow. With an increase
in the strength of
flow, a feedback from the elastic force which tends to preserve the previous relaxation,
results in an enhanced tumbling time. Thus, resulting in a decrease in the tumbling frequency 
$\nu\tau_0$ for small Wi. 
On the other hand, for strong flow, when Wi is quite large, tumbling occurs more frequently 
due to the increase in the rotational component of shear force dominating the tumbling dynamics. 
Consequently, for some intermediate value of Wi, the tumbling frequency $\nu\tau_0$ 
hits a minima and our numerical calculations also corroborate this fact.

We also learn from Fig.~\ref{fig5} that the tumbling time $\nu\tau_0$ changes significantly
at any given value of Wi even for slight decrease in the value of $\alpha$.
This is because the fluctuations in the relative coordinate are reduced for low values 
of $\alpha$, thus making tumbling a flow dominant phenomena which tends 
to keep the dumbbell oriented along the flow, hence the increase in tumbling time. 
Interestingly, for viscoelastic case, the increase in $\nu$ starts at 
quite a high value of Wi as compared to the viscous case. This shift is consequent of 
the prolonged relaxation in viscoelastic medium due to inherent elasticity.

Now, in the limit of very large values of the Weissenberg number, \textit{i.e.}- for
Wi$\gtrapprox 100$ for which $\nu\tau_0$ is observed to monotonically increase with Wi
with exponent close to 0.67 ($\alpha = 1$) as $\alpha$ approaches unity. However, for
arbitrary values of $\alpha$ we have not been able to extract reliable statistics of
tumbling times so as to furnish a value of the growth exponent defining the scaling
behavior: $\nu\tau_0 \approx \text{Wi}^{\mu(\alpha)}$ \cite{supp}. This is because with decreasing
$\alpha$, a tumbling event becomes rare to observe making it computationally very challenging
to record them in a finite amount of computational time available to us.


\textit{Conclusion}:
Viscoelasticity is more of a rule rather than exception, and motivated by this, we have
studied in this paper the transport and tumbling properties of polymers in a viscoelastic
fluid under shear. Using dumbbells as representative, we provide analytical results for
the motion of center of mass and separation between the two masses. For the simplest
case of a harmonic spring connecting the two masses, we that the mean square displacement
of the center of mass follows: $\langle x^2_c(t) \rangle \sim \dot{\gamma}^2
t^{\alpha+2}, ~0< \alpha <1$, generalizing the earlier result: $\langle x^2_c(t) \rangle
\sim \dot{\gamma}^2t^3 ~(\alpha = 1)$. On the other hand, fluctuations in the relative
coordinate also grow monotonically with time, with $\langle x^2_r(t) \rangle \sim t^\beta$,
where $\beta = 2(1-\alpha)$ up to $\alpha \approx 0.25$ and approaches 0 as $\alpha$
approaches unity. Consequently, the system of two masses connected by a harmonic spring
does not achieve a steady-state. In other words, the extensively studied Rouse polymer
is inappropriate to address the dynamics of polymers in viscoelastic medium under
shear. We remedy this pathology by introducing a nonlinear
spring in the form of FENE-LJ interaction which restricts the separation of the two
masses to a maximum allowed limit. Employing the nonlinearity in the system we address
tumbling of dumbbells and find that the effect of viscoelasticity in medium is to
slow down the characteristic tumbling frequency at finite Wi. We hope that our work
motivates further studies along this direction, particularly the effect of hydrodynamic
interactions on tumbling aspects.

\begin{acknowledgements}
We thank Dibyendu Das for insightful discussions. SS and SK also acknowledge the
financial assistance from SERB and INSPIRE program of DST, New Delhi, India.
\end{acknowledgements}

\end{document}


\title{Supplementary Information for Tumbling dynamics of a dumbbell
in viscoelastic shear flow}
\author{Sadhana Singh}
\affiliation{Department of Physics, Banaras Hindu University, Varanasi-221005}
\author{R. K. Singh}
\email[]{rksinghmp@gmail.com}
\affiliation{Department of Physics, Banaras Hindu University, Varanasi-221005}
\affiliation{Department of Physics, Indian Institute of Technology, Bombay, Powai,
Mumbai-400076}
\author{Sanjay Kumar}
\affiliation{Department of Physics, Banaras Hindu University, Varanasi-221005}

\maketitle

\section{Extension to polymers}
\begin{figure}
\includegraphics[width=0.45\textwidth]{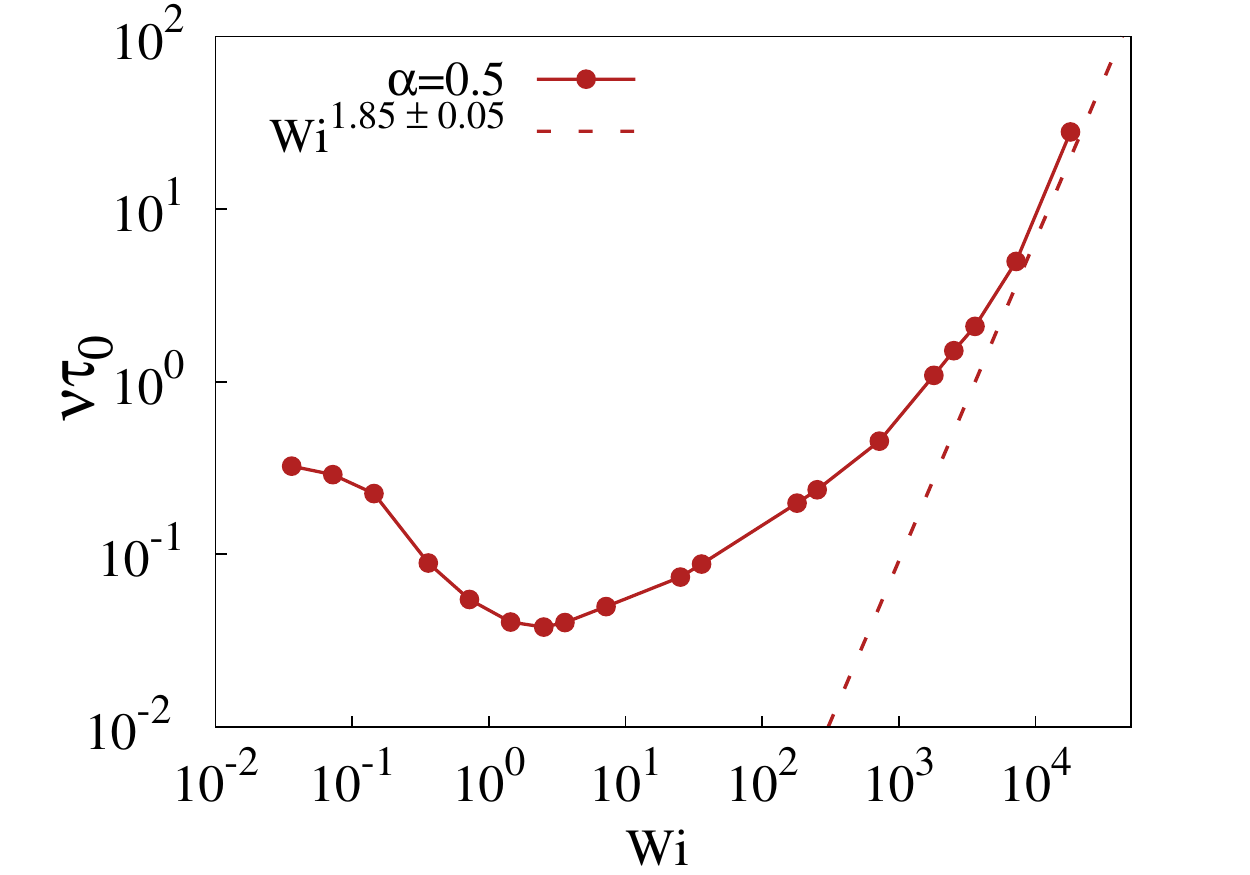}
\caption{Limiting behavior of the tumbling frequency $\nu\tau_0$ with Weissenberg
number Wi for a polymer of size $N = 20$ for $\alpha = 1/2$. For Wi$\gg 1$, we find
that $\nu\tau_0\sim \text{Wi}^{1.85\pm 0.05}$.}
\label{fig6}
\end{figure}

Let us generalize to the case of polymers in viscoelastic shear flow.
We take a polymer of size $N=20$ with $\alpha = 1/2$. The equation
of motion for the $i$-th monomer reads:
\begin{align}
\label{polymer}
\ddot{\bf r}_i + \int^t_0 dt' \eta(t-t')(\dot{\bf r}_i-\dot{\gamma}y_i{\bf i})(t')
= -\sum_{j\neq i}\nabla_i V(|{\bf r}_i - {\bf r}_j|) + \dot{\gamma}\dot{y}_i {\bf i} +
{\bf \xi}_i(t),
\end{align}
where $V=V_{FENE}+V_{LJ}$ and $\xi_i$ is an $N$-tuple of Gaussian random variables
with power law correlation, with only the repulsive part of the LJ potential (athermal good solvent
conditions) employed to address excluded volume effect amongst non-bonded beads.

Even though the polymer is an extended object composed of $N$ monomers,
its center of mass ${\bf r}_c = \sum_i {\bf r}_i/N$ moves like a free particle, with its
mean square displacement $\langle {\bf r}^2_c(t) \rangle \sim t^{\alpha + 2}$, as
shown previously for dumbbells. The diffusion coefficient for the center of mass motion is,
however, reduced by a factor of $N$. In order to calculate the tumbling time of the polymer for
given value of Weissenberg number Wi, we numerically solve Eqns.~(\ref{polymer}) and
calculate the zero-crossing times of the end-to-end vector $x_1-x_N$ along the shear
flow. We find that for Wi $\gg 1$,
the characteristic tumbling frequency $\nu\tau_0 \sim \text{Wi}^{\mu}$ with
$\mu \approx 1.85\pm 0.05$ for $\alpha = 1/2$ (\textit{cf.} Fig.~\ref{fig6}),
where $\tau_0$ is the characteristic relaxation
time of the autocorrelation of $x_1-x_N$ in the absence of shear. This compares well,
for example, with the corresponding exponent for nonlinear dumbbells, for which
$\nu\tau_0 \approx \text{Wi}^{1.9}$ for $\alpha = 1/2$.

An important thing
to notice about polymer tumbling is that the exponent characterizing tumbling in viscoelastic
shear flow is identical in value (within errorbars) to the case of dumbbell. However,
there is an important difference between the $\nu\tau_0$ vs Wi graphs for the polymer
as opposed to the dumbbell. It is evidently clear from Fig. 5 (main text) and Fig.~\ref{fig6}
is that location of minima for polymer is about two orders of magnitude less than the
corresponding value for dumbbells. This is consequent of the fact that a polymer, being
an extended object, will exhibit a tumbling event even in the extreme case when the
two ends of polymer are close to each other (like in a U-shape). On the other hand, a
dumbbell mostly remains as \textit{extended} thus requiring more time to tumble.

In addition, as we have previously seen \cite{sdhn}, polymers in a purely
viscous medium do not exhibit a minima in the $\nu\tau_0$ vs Wi curve, in a good
solvent. However, exhibition of a minima for a polymer tumbling in a viscoelastic shear
flow implies that the elastic component of the background fluid tends to slow down
the occurrence of tumbling. At high shear rates, however, the behavior is similar to
motion in viscous media.

\section{Embedding for overdamped motion in shear flow}
For simplicity, we discuss the concept for a single particle moving in the $x-y$
plane in a viscoelastic fluid experiencing shear along the $x$-axis. The generalized
Langevin equation describing the overdamped dynamics reads:
\begin{subequations}
\label{gle_sh1}
\begin{align}
&\int^t_0 dt' \eta(t-t')(\dot{x}-\dot{\gamma}y)(t') = -V'_x(x,y) + \xi_x(t),\\
&\int^t_0 dt' \eta(t-t') \dot{y}(t') = -V'_y(x,y) + \xi_y(t),
\end{align}
\end{subequations}
where $\xi_{x,y}(t)$ are Gaussian random variables with correlations
\begin{align}
\langle \xi_x(t) \xi_y(t') \rangle = \delta_{xy} k_BT \eta(|t-t'|).
\end{align}
The friction kernel is a power-law decaying function of time: $\eta(t) =
\eta_\alpha t^{-\alpha}/\Gamma(1-\alpha)$ with $0 < \alpha < 1$. In order to solve
Eq.~(\ref{gle_sh}), we employ the technique of Markovian embedding. We follow the
review by Goychuk \cite{advchemphys} and outline the methodology here for motion
in shear flow. As only the $x$ coordinate involves a contribution from shear flow,
we focus only on motion along $x$-direction. Define $u_i = -k_i(x-\dot{\gamma}
\int^t_0 dt' y(t') -x_i)$, where $x_i$ are auxiliary variables following
\begin{align}
\label{aux}
\eta_i\dot{x}_i = k_iu_i + \sqrt{2\eta_ik_BT}\xi_i(t),
\end{align}
with $k_i = C_\alpha(b)\eta_\alpha \nu^\alpha_0/[b^{\alpha(i-1)}\Gamma(1-\alpha)]$ and
$\eta_i = C_\alpha(b) \eta_\alpha \nu^{\alpha-1}_0 b^{(1-\alpha)(i-1)}/\Gamma(1-\alpha)$,
(\textit{cf.} Eq.~(23) in \cite{advchemphys}). The numbers $k_i$ and $\eta_i$ define the
Ornstein-Uhlenbeck processes
\begin{align}
\label{zeta_i}
\dot{\zeta}_{i,x}(t) = -\nu_i \zeta_{i,x}(t) + \sqrt{2k_i\nu_ik_BT}\xi_{i,x}(t),
\end{align}
and are useful in approximating the power-law decay form with a sum of exponentials as
$\eta(t) = \sum_i k_i e^{-\nu_i t}$, where $\nu_i = k_i/\eta_i$. The idea of representing
power-law decay with such
a form is fairly old \cite{palmer}. The equivalence of Eq.~(\ref{gle_sh1}) and (\ref{aux})
is easily shown, thus providing a way to numerically solve the former.

\section{Markovian embedding for underdamped motion in shear flow}
The generalized Langevin equation describing the dynamics of an underdamped particle
in the $x-y$ plane with a shear flow along $x$-axis reads:
\begin{subequations}
\label{gle_sh}
\begin{align}
&\ddot{x} + \int^t_0 dt' \eta(t-t')(\dot{x}-\dot{\gamma}y)(t') = -V'_x(x,y) + 
\dot{\gamma}\dot{y} + \xi_x(t),\\
&\ddot{y} + \int^t_0 dt' \eta(t-t') \dot{y}(t') = -V'_y(x,y) + \xi_y(t),
\end{align}
\end{subequations}
where the symbols retain their usual meaning. The Markovian 
embedded form for the above equation in terms of auxiliary variables ${\{u_{i,x}\}^N_{i=1}}$ 
and ${\{u_{i,y}\}^N_{i=1}}$ is:
\begin{subequations}
\label{markov}
\begin{align}
\dot{x} &= v_x,\\
\dot{v}_x &= -V_x'(x,y) + \dot{\gamma}\dot{y} + \sum^N_{i=1} u_{i,x}(t),\\
\dot{u}_{i,x} &= -k_i(v_x-\dot{\gamma}y) - \nu_i u_{i,x} + \sqrt{2 \nu_i 
k_i k_BT} \xi_{i,x}(t),\\
\dot{y} &= v_y,\\
\dot{v}_y &= -V_y'(x,y) + \sum^N_{i=1} u_{i,y}(t),\\
\dot{u}_{i,y} &= -k_iv_y - \nu_i u_{i,y} + \sqrt{2 \nu_i 
k_i k_BT} \xi_{i,y}(t),
\end{align}
\end{subequations}
where $\xi_{i,x}$ and $\xi_{i,y}$ are Gaussian white noise variables with mean 
zero and with correlations: $\langle \xi_{i,x}(t) \xi_{j,y}(t')\rangle = \delta_{ij} 
\delta_{xy} \delta(t-t')$. Now, from Eq.~(A.3c) we have:
\begin{align*}
u_{i,x}(t) = -\int^t_0 dt'(v_x-\dot{\gamma}y)(t')k_i e^{-\nu_i(t-t')} + 
\int^t_0 dt' \sqrt{2\nu_i k_i k_BT}\xi_{i,x}(t')e^{-\nu_i(t-t')},
\end{align*}
wherein we have used the initial conditions $u_{i,x}(0)=0$. Substituting this in 
(A.3b) we have
\begin{align*}
\dot{v}_x &= -V'_x(x,y) + \dot{\gamma}\dot{y} -\int^t_0 dt'(v_x-\dot{\gamma}y)(t')
\sum^N_{i=1} k_i e^{-\nu_i(t-t')}
 + \sum^N_{i=1}\int^t_0 dt' \sqrt{2\nu_i k_i k_BT}\xi_{i,x}(t')e^{-\nu_i(t-t')},\\
&= -V'_x(x,y) + \dot{\gamma}\dot{y} - \int^t_0 dt' (v_x-\dot{\gamma}y)(t') 
\eta(t-t') + \xi_x(t),
\end{align*}
where to obtain the last step we have used the solution of Eq. (A.4) and 
$\xi_x(t) = \sum_i \zeta_{i,x}(t)$. Similar calculations for the $y$-coordinate show 
the equivalence of Eq.~(\ref{gle_sh}) to the Markov embedded form (\ref{markov}).
An important difference to be noted from the case of embedded form for overdamped
motion is the absence of memory term for the underdamped dynamics.

It is noted that the representation of a power-law decaying function with a sum of 
exponentials serves a good approximation to the former only in a finite range 
$[t_i, t_f]$, beyond which there are exponential cutoffs. For all practical purposes, 
like the one considered here, the range $[t_i, t_f]$ is sufficient. Following 
Ref.~\cite{advchemphys}, we use an $N=16$ term exponential approximation for the 
power-law decaying memory kernel employing decade scaling $b = 10$. Also, the fastest 
time-scale $\nu_0 = 10^3$.